\documentclass{ar-1col-S2O}
\usepackage[utf8]{inputenc}
\DeclareUnicodeCharacter{2009}{\,}
\usepackage[square, numbers]{natbib}
\usepackage{url}
\usepackage{amsmath}
\usepackage{soul}

\usepackage[normalem]{ulem}
\usepackage[dvipsnames,x11names]{xcolor}
\usepackage[colorlinks=false, citecolor=black, urlcolor=purple, linkcolor=purple]{hyperref}
\usepackage{graphicx}

\jname{Xxxx. Xxx. Xxx. Xxx.}
\jvol{AA}
\jyear{YYYY}
\doi{10.1146/((please add article doi))}

\begin{document}

\markboth{Keidar et al.}{Resetting: Prediction, Inference, Design}

\title{Stochastic Resetting: A Non-Equilibrium Framework for Prediction, Inference and Design}

\author{Tommer D. Keidar,$^1$ Sagi Meir,$^1$, Nir Sherf,$^1$ Rémi Goerlich,$^1$, Shlomi Reuveni,$^{1,2,4}$, Yael Roichman,$^{1,2,3}$, Barak Hirshberg,$^{1,2,4}$
\affil{$^1$School of Chemistry, Tel Aviv University, Tel Aviv 6997801, Israel.}
\affil{$^2$The Center for Physics and Chemistry of Living Systems, Tel Aviv University, Tel Aviv 6997801, Israel.}
\affil{$^3$School of Physics and Astronomy, Tel Aviv University, Tel Aviv 6997801, Israel.}
\affil{$^4$The Center for Computational Molecular and Materials Science, Tel Aviv University, Tel Aviv 6997801, Israel.; Emails: shlomire@tauex.tau.ac.il, roichman@tauex.tau.ac.il, hirshb@tauex.tau.ac.il}}

\begin{abstract}
Stochastic resetting has evolved from a simple model of diffusive search acceleration into a general framework for predicting, inferring, and controlling stochastic dynamics far from equilibrium. Its defining features, i.e., the creation of non-equilibrium steady states and the acceleration of first-passage kinetics, are increasingly relevant across physical chemistry, from biological restart mechanisms to molecular simulations and colloidal experiments. We review the renewal theory underlying stochastic resetting and show how it enables prediction of reset dynamics from properties of the underlying process, while also allowing the latter to be inferred from the resetting-accelerated dynamics. We then discuss applications to state preparation, enhanced sampling, kinetic inference, and training and sampling of machine learning models. Finally, we review recent advances in adaptive resetting, environmental feedback, many-body dynamics, and thermodynamic costs of resetting. These developments establish
new opportunities for controlling stochastic dynamics with resetting across theory, simulations, and experiments.
\end{abstract}

\maketitle

\section{INTRODUCTION}
\subsection{What is resetting?}
Stochastic resetting is the procedure of stopping a random process and re-initializing it. Why would anyone want to restart a stochastic process? The pioneering work of Evans and Majumdar \cite{Evans_diffusion_2011} gave two compelling reasons, by investigating the effect of resetting on diffusion.
Without resetting, the mean time it takes a diffusing particle to reach a target for the first time, i.e., the mean first-passage time (FPT) \cite{redner2001guide,metzler2014first,bray2013persistence}, diverges.
Introducing resetting at any positive, finite rate renders the mean FPT finite, and an optimal rate minimizes it. The intuition is simple: resetting too often prevents the particle from moving, and resetting too slowly gives back regular diffusion; in both limits the mean FPT diverges.
Resetting also fundamentally alters the long-time behavior: without it, the particle density expands indefinitely, whereas with resetting it reaches a time-independent, non-equilibrium steady state.
Again the intuition is simple: resetting confines the particle to a region whose size is set by the distance it diffuses between resets.
These two observations, that resetting can optimize mean FPTs and generate non-equilibrium steady states, have since been extended to stochastic processes far beyond the toy model of diffusion, establishing a new field of research.

\subsection{Why should a physical chemist care?}

\begin{figure}
    \centering
    \includegraphics[width=\linewidth]{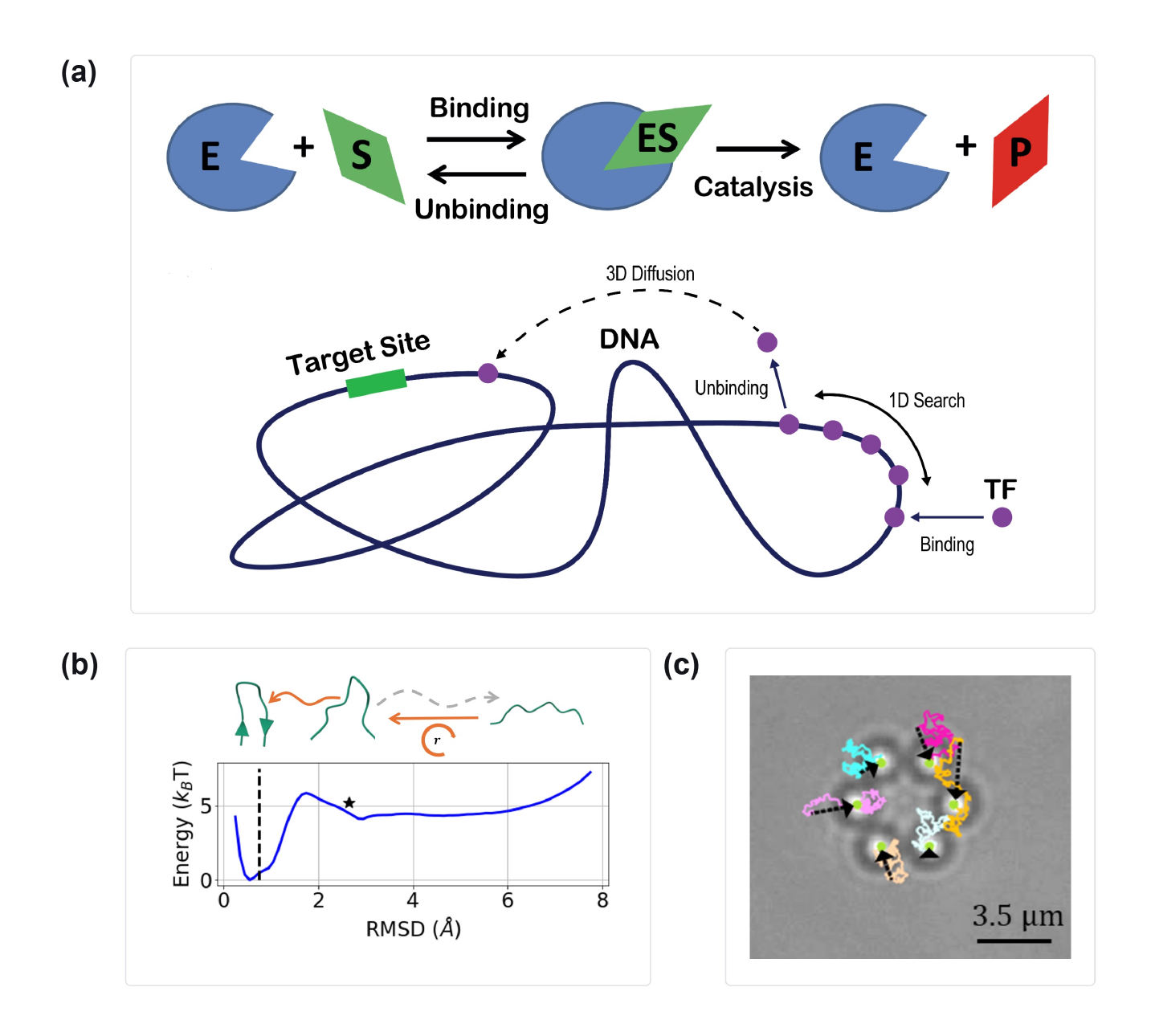}
    \caption{\textbf{Why should a physical chemist care about stochastic resetting?} (a) Resetting arises naturally in chemical and biological systems, e.g., in the Michaelis-Menten reaction scheme and in facilitated diffusion of proteins along the DNA. (b) Resetting is a powerful tool in computational chemistry, e.g., in sampling rare conformational transitions of biomolecules. (c) Resetting is a proving ground for experimental non-equilibrium statistical mechanics, e.g., six $1.5~\mu m$ silica colloidal particles diffusing in water are reset using holographic optical tweezers.}
    \label{fig:reasons}
\end{figure}

    Stochastic resetting has found widespread applications including search processes \cite{Evans_diffusion_2011, Pal_first_2017, Chechkin_random_2018}, queuing systems \cite{bonomo2022mitigating, bonomo2025queues, roy2024queues},  and randomized computer algorithms \cite{luby1993optimal, alt1996method, langville2004deeper, Avrachenkov2010improving}.
But what is the motivation for a physical chemist to be interested in resetting? 
We argue the answer is three-fold (see Fig.~\ref{fig:reasons}):
\begin{enumerate}
    \item \textbf{Resetting arises naturally in chemical and biological systems.} Many chemical and biological processes are repeatedly interrupted and restarted by intrinsic mechanisms. In enzymatic reactions, unproductive substrate unbinding resets the catalytic cycle~\cite{Reuveni_Role_2014, Scheerer2025Interplay}. During facilitated diffusion, transcription factors (TF) repeatedly abandon one-dimensional searches on DNA and restart them elsewhere following three-dimensional diffusion in the bulk \cite{halford2004site, jangid2026dna}. Molecular chaperones rescue misfolded proteins by unfolding them, giving them another chance to reach the native state \cite{thirumalai2020iterative, Pal2023}. Similar mechanisms appear in animal search and foraging, where repeated returns to a home base improve search efficiency \cite{papi2012animal, paramanick2024uncovering, Mercado-Vasquez_2018Lotka}.
    
    \item \textbf{Resetting is a useful tool in computational chemistry~\cite{Blumer2025review}.} For example, it has been used for enhanced sampling, either standalone~\cite{blumer_stochastic_2022} or combined with methods such as Metadynamics~\cite{blumer2024combining}.  It has accelerated simulations of protein conformational changes in explicit solvent~\cite{blumer2024combining, Church2025} and improved kinetic inference from Metadynamics~\cite{blumer2024short}. More broadly, resetting can be viewed as a birth-death algorithm~\cite{Pampel2023Sampling} that selectively terminates and reinitializes trajectories, placing it among trajectory-based rare-event methods that terminate and re-spawn trajectories, such as milestoning~\cite{Elber2020Milestoning, Elber2021Modeling} and weighted-ensemble sampling~\cite{Zhang2010weighted}.

    \item \textbf{Resetting is a proving ground for non-equilibrium thermodynamics.} Resetting steers a system out of equilibrium through stochastic perturbation rather than deterministic driving or intrinsic activity. This conceptually simple mechanism produces core non-equilibrium phenomena: persistent probability fluxes that violate detailed balance, non-equilibrium steady states, and dynamic transitions in survival statistics \cite{fuchs_stochastic_2016, Eule_2016, Pal_first_2017}.
    It is also often analytically tractable, allowing exact calculation of entropy production rates, thermodynamic efficiency bounds, and optimal control protocols \cite{pal_integral_2017, Pal2021, goerlich_time-energy_2026}. Recent colloidal experiments rigorously validate these predictions, establishing resetting as a premier framework where non-equilibrium theory and experiment align \cite{tal-friedman_experimental_2020, besga_optimal_2020}.

\end{enumerate}

\subsection{Disclaimer}
The primary goal of this paper is not to provide a comprehensive review of stochastic resetting and its applications, as several excellent reviews and perspectives are already available~\cite{Evans_Stochastic_2020,gupta_stochastic_2022, pal2022inspection, kundu2024preface}.
Instead, we introduce stochastic resetting to the physical chemistry community as a framework for prediction, inference, and design.  Resetting can provide useful tools for physical chemistry, while applications in colloidal, biological, and chemical systems can motivate new developments in resetting theory. This reciprocal exchange has already begun \cite{paramanick2024uncovering, Church2025, tal-friedman_experimental_2020, besga_optimal_2020, altshuler_environmental_2024, Pal2020Search,   keidar_adaptive_2025, biroli_extreme_2023, biroli2025experimental, vatash_manybody_2025, vatash_numerical_2025, Sokolov2023linear}, as will be discussed throughout this review. Accordingly, we emphasize developments especially relevant to physical chemistry, including recent advances not covered in earlier reviews.
In each, we highlight the interplay between theory, experiments, and simulations. 
Finally, we focus on resetting in classical systems. We do not address the emerging field of resetting in quantum systems, which merits its own review \cite{Mukherjee2018quantum, Yin2023restart, Kulkarni2023Generating, Yin2025restart,sevilla2023dynamics,dubey2023quantum,perfetto2021designing,turkeshi2022entanglement}.

\subsection{Structure of the review}

We begin with the theory of stochastic resetting, emphasizing the universal framework underlying both steady-state and kinetic (first-passage) phenomena. We then explore resetting as a practical tool for acceleration and inference in molecular simulations and machine learning. Next, we discuss the growing role of information and feedback, which extends the theory to adaptive, state-dependent protocols. We then show how resetting can be used to design the properties of the process undergoing resetting, e.g., steady states and transitions between them. Expanding to collective phenomena and many-body resetting, we examine how multi-particle interactions and resetting protocols, such as global and local resetting, induce non-trivial spatial correlations. Finally, we discuss the thermodynamics of resetting and the energetic costs of maintaining steady states, accelerating searches, and driving rapid transitions, before concluding with an outlook on open challenges relevant to physical chemistry.

\section{THEORY OF STOCHASTIC RESETTING}
\label{sec:theory}

Many processes in physical chemistry are random in nature: a protein undergoing a conformational change, an enzyme catalyzing a chemical reaction, or a molecule diffusing on the surface of a metal.
Whether resetting occurs naturally, is used to accelerate a simulation, or is applied experimentally to drive a system out of equilibrium, we would like to predict its effects on steady-state properties and kinetics: how much resetting accelerates transitions, and which non-equilibrium steady states replace the Boltzmann distribution. Below, we summarize the theoretical framework needed to address these questions.

\begin{figure}[t]
\includegraphics[width=0.75\linewidth]{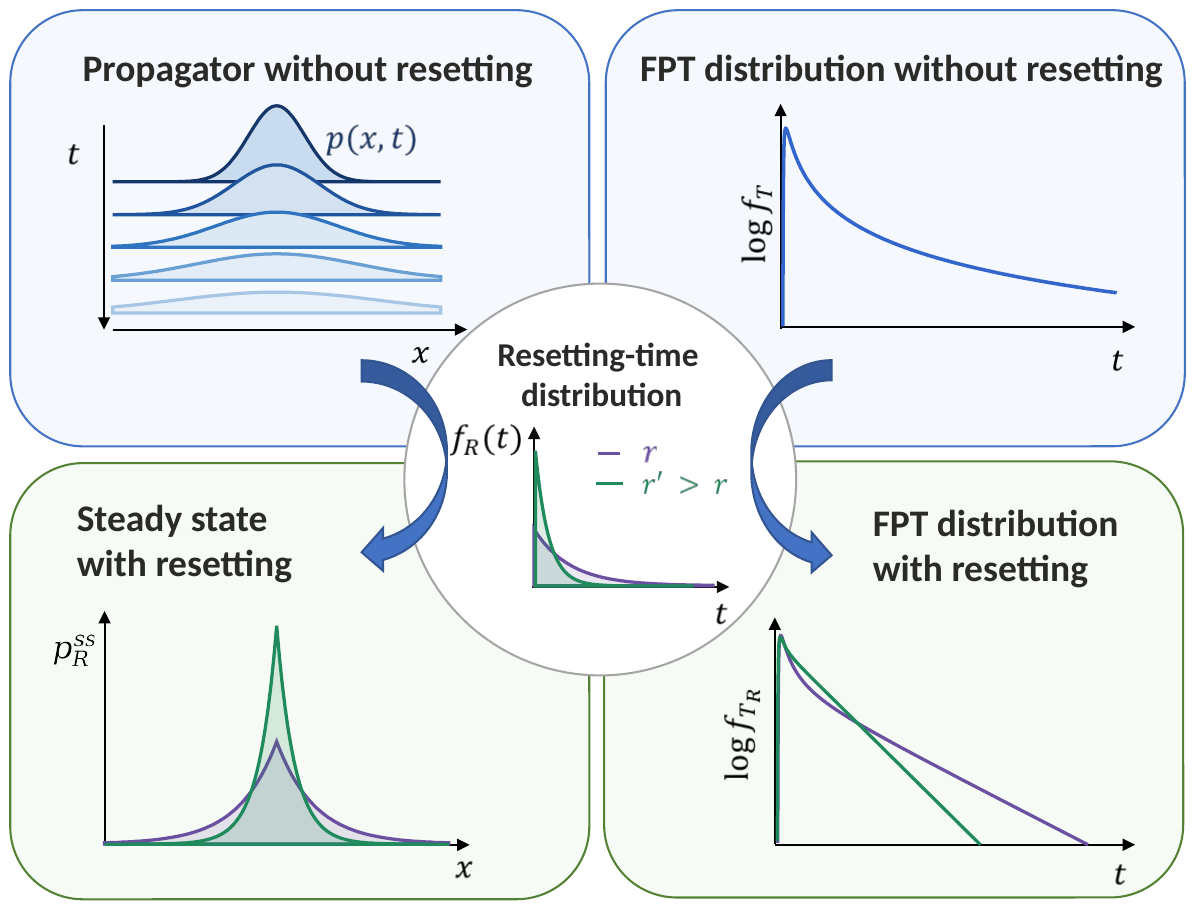}
\caption{\textbf{Predicting reset dynamics from quantities without resetting.} The propagator without resetting $p(\boldsymbol{x},t)$ and first-passage-time distribution $f_T(t)$ serve as independent inputs that, when combined with the resetting-time distribution $f_R(t)$ via Eqs.~\ref{eq:propagator_renewal_Laplace} and \ref{eq:MFPT_general_resetting}, completely determine the corresponding observables under resetting. Illustrated here for the case of free diffusion with poison resetting with rate $r$, but applicable to general stochastic processes.}
\label{fig2}
\end{figure}

Imagine a system undergoing some stochastic evolution. Without resetting, we denote the probability of finding the system at state $\boldsymbol{x}$ at time $t$ by $p(\boldsymbol{x},t)$. We then reset the system at times taken from some distribution with density $f_R(t)$, where $R$ is the random variable for the time between resetting events, often called the resetting timer. We would like to know $p_R(\boldsymbol{x},t)$, the probability that the process undergoing resetting is in state $\boldsymbol{x}$ at time $t$. There are two contributions to this probability:
\begin{enumerate}
    \item The process was not reset until time $t$, in which case the probability of finding it at state $\boldsymbol{x}$ is $p(\boldsymbol{x},t)$.
    \item The process was reset at least once, and the first resetting event occurred at time $\tau\in[0,t]$. In that case, the probability of finding it at state $\boldsymbol{x}$ is given by $p_R(\boldsymbol{x},t-\tau)$.
\end{enumerate}
Therefore, $p_R(\boldsymbol{x},t)$ has two contributions, one from the dynamics without resetting, weighted by the probability that no reset occurred until time $t$, and one from the dynamics with resetting, weighted by the probability of resetting occurring at time $\tau$ and integrated over all $\tau$. This yields the celebrated renewal equation~\cite{Evans_Stochastic_2020, Pal_diffusion_2016, Gupta_Fluctuating_2014}:
\begin{equation}\label{eq:propagator_renewal}
    p_R(\boldsymbol{x},t)=p(\boldsymbol{x},t)\Psi_R(t)+\int_0^t f_R(\tau)p_R(\boldsymbol{x},t-\tau)\,d\tau.
\end{equation}
In Equation~\ref{eq:propagator_renewal}, $\Psi_R(t)\equiv\text{Pr}(R > t)=\int_t^\infty f_R(\tau)\,\mathrm{d}\tau$ is the probability that no resetting event happened until time $t$. 
Since the second term is a convolution, Equation (\ref{eq:propagator_renewal}) can be solved by applying a Laplace transform
\begin{equation}\label{eq:propagator_renewal_Laplace}
    \tilde{p}_R(\boldsymbol{x},s)=\frac{\mathcal{L}\{p(\boldsymbol{x},t)\Psi_R(t)\}(s)}{1-\tilde{f}_R(s)},
\end{equation}
where we denote the Laplace transform of some function $g(t)$ by $ \tilde{g}(s) \equiv \mathcal{L}\{g(t)\}$.
This general and powerful result is illustrated in Fig.~\ref{fig2}: given $p(\boldsymbol{x},t)$ and $f_R(t)$, it provides a direct means to predict $p_R(\boldsymbol{x},t)$. We stress that $p(\boldsymbol{x},t)$ can be obtained in any way: exact solutions, simulations, or experiments. A particularly useful case is resetting at a constant rate $r$, where $f_R(t)=re^{-rt}$, often called Poisson resetting. 
In this case, one finds~\cite{Evans_diffusion_2011}
\begin{equation}\label{eq:poisson_reset_time_evoultion}
    \tilde{p}_R(\boldsymbol{x},s)=\frac{r+s}{s}\tilde{p}(\boldsymbol{x},r+s).
\end{equation}

Systems undergoing stochastic resetting typically converge to a non-equilibrium steady-state, regardless of whether the underlying process reaches steady-state. According to the final value theorem \cite{schiff1999laplace}, the steady-state distribution is given by $p_R^{ss}(x)\equiv\lim_{t\to\infty}[ {p}_R(x,t) ]=\lim_{s\to0^+} [ s\tilde{p}_R(x,s) ]$, provided this limit exists.
For Poisson resetting, where the mean time between resetting evens is $r^{-1}$, one finds~\cite{Evans_diffusion_2011}
\begin{equation}\label{eq:poisson_reset_time_evoultion_NESS}
    p_R^{ss}(\boldsymbol{x})=r\tilde{p}(\boldsymbol{x},r).
\end{equation}

After discussing the time evolution and steady-state, we now look at kinetics under resetting.
Consider a stochastic process, as before, but now we have a well-defined target state and measure the first time of reaching it, the first-passage time (FPT). Since the process is stochastic, the FPT is distributed according to some first-passage density $f_T(t)$, where $T$ is the random variable representing the FPT. Denoting with $T_R$ the FPT under resetting with a timer $R$, we look to find its distribution $f_{T_R}(t)$. Employing the same arguments that led to Eq.~\ref{eq:propagator_renewal}, one gets a renewal equation for the first-passage time distribution~\cite{Pal_first_2017,Chechkin_random_2018}
\begin{equation}
    f_{T_R}(t)=f_T(t)\Psi_R(t)+\int_0^t f_R(\tau)\Psi_T(\tau)f_{T_R}(t-\tau)\,\mathrm{d}\tau,
\end{equation}
where $\Psi_T(t)\equiv\Pr(T> t)=\int_t^{\infty}f_T(\tau)\,\mathrm{d}t$ is the probability that, for the process without resetting, first-passage did not occur up to time $t$. It appears in the integrand because for the first resetting event to happen at time $\tau$, we must also require that the process did not complete up to that time.

This equation can be solved by taking the Laplace transform, as in Eq.~\ref{eq:propagator_renewal_Laplace}, to get
\begin{equation}
    \tilde{f}_{T_R}(s)=\frac{\mathcal{L}\{f_T(t)\Psi_R(t)\}(s)}{1-\mathcal{L}\{f_R(t)\Psi_T(t)\}(s)}.
\end{equation}
For the particular case of Poisson resetting at rate $r$, one has \cite{Reuveni_optimal_2016}
\begin{equation}\label{eq:poisson_reset_FPT_MFPT}
    \tilde{f}_{T_R}(s)=\frac{(r+s)\tilde{f}_T(s+r)}{s+r\tilde{f}_T(s+r)}.
\end{equation}
From the Laplace transform we can obtain all the moments of the first-passage time distribution by taking derivatives with respect to $s$, and evaluating at $s\to0$. For example, the mean FPT is
\begin{equation}\label{eq:MFPT_general_resetting}
    \langle T_R\rangle=\frac{\int_0^\infty \Psi_T(t)\Psi_R(t)\,\mathrm{d}t}{\int_0^\infty f_T(t)\Psi_R(t)\,\mathrm{d}t}=\frac{\langle\min(T,R)\rangle}{\Pr(T\leq R)},
\end{equation}
where $\min(T,R)$ is the minimal value between the resetting timer and the FPT without resetting~\cite{Pal_first_2017, Chechkin_random_2018}. As illustrated in Fig.~\ref{fig2}, Eq.~\ref{eq:MFPT_general_resetting} shows that if we know the FPT distribution without resetting, we can predict the mean FPT with resetting for any resetting timer $R$. This is a very useful result: it determines whether, and by how much, resetting lowers the mean FPT. For Poisson resetting at rate $r$, one obtains~\cite{Reuveni_optimal_2016}
\begin{equation}\label{eq:MFPT_Poisson_resetting}
    \langle T_R\rangle=\frac{1-\tilde{f}_T(r)}{r\tilde{f}_T(r)}.
\end{equation}
Taylor expanding $\langle T_R\rangle$ as a function of $r$ for small $r$, and using $\tilde{f}_T''(0)=\langle T^2\rangle$ gives~\cite{Reuveni_Role_2014, Rotbart2015Michaelis}
\begin{equation}
    \langle T_R\rangle=\langle T\rangle+\frac{r}{2}\left(\langle T\rangle^2-\sigma^2(T)\right)+o(r),
\end{equation}
where $\sigma(T)$ is the standard deviation of the FPT without resetting. Examining this expansion, while keeping in mind that $r>0$, one obtains a sufficient condition for Poisson resetting to expedite an FPT process,
\cite{Reuveni_Role_2014, Rotbart2015Michaelis}
\begin{equation}
    \sigma(T)>\langle T\rangle.
\end{equation}
Namely, if the FPT distribution without resetting is broad enough, Poisson resetting expedites the process. If the mean or standard deviation of the FPT without resetting diverges, a separate analysis shows that a small resetting rate is guaranteed to expedite it \cite{Reuveni_Role_2014}.

\section{RESETTING AS A TOOL FOR ACCELERATION AND INFERENCE}
The theory just reviewed predicts the steady-state, dynamics, and first-passage properties of a process undergoing resetting from observables of the process without resetting and the resetting-time distribution. We now show how it accelerates: 1) relaxation to steady state, 2) rare transitions in molecular simulations, and 3) the training of machine-learning models and sampling from them.
In the latter two, the properties of the process without resetting can also be inferred efficiently from the resetting-accelerated process.
This makes resetting a useful tool for enhanced sampling in both molecular simulations and machine learning models.

\subsection{Accelerating relaxation to steady state}
\label{sec:StateToState}

In physical chemistry and molecular simulations, slow state preparation is often a limiting factor. Preparing a target macromolecular conformation, chemical state, or ensemble requires relaxation of the full probability distribution, not just convergence of a single mean observable, and is commonly quantified by the Kullback--Leibler divergence or the total variation distance. Long relaxation times often arise from metastable configurations, high activation barriers, or diffusive bottlenecks that delay probability redistribution across phase space.

State-to-state transitions are commonly accelerated by deterministic control. Shortcut-to-equilibration protocols, such as Engineered Swift Equilibration, prepare a desired final distribution faster than ordinary thermal relaxation by applying a time-dependent potential or force field~\cite{martinez_engineered_2016, patra_shortcuts_2017, guery-odelin_shortcuts_2019}, showing that relaxation accelerates when the probability-redistribution pathway is actively controlled rather than left to thermal dynamics.

Stochastic resetting provides a different type of control. Instead of continuously reshaping the potential, it intermittently removes probability from the system's current state and re-injects it into a prescribed state or region. For free diffusion, stochastic resetting converts an unconfined transient process into a localized non-equilibrium steady state~\cite{Evans_diffusion_2011, Evans_Stochastic_2020}. The approach to this steady state is spatially non-uniform: probability near the reset location relaxes rapidly, whereas distant regions converge more slowly through an outwardly ballistically propagating relaxation front~\cite{majumdar_dynamical_2015}. In a potential energy landscape, this re-injection competes with potential-driven thermal relaxation~\cite{Pal_diffusion_2016, singh_resetting_2020}, so relaxation kinetics depend on both the landscape and the resetting-time statistics~\cite{singh_general_2022}.

Goerlich et al. \cite{goerlich_resetting_2024} suggested using this redistribution of probability as a stochastic state-to-state driver. They noted that Poisson resetting of a freely diffusing particle generates a Laplace steady state, identical to the Boltzmann equilibrium of a particle in a V-shaped potential for an appropriate resetting rate~\cite{Evans_diffusion_2011, goerlich_resetting_2024, Evans_2013Optimal}, which motivated the following transition-control protocol. The system starts in the equilibrium of a V-shaped potential; the potential is then switched off and Poisson resetting applied. When the evolving distribution reaches the Boltzmann distribution of a new V-shaped target potential, resetting is turned off and the target potential switched on, leaving the system already at the equilibrium of the new potential.

This protocol always accelerates the transition relative to a reference protocol in which the potential is changed abruptly, and the system relaxes thermally~\cite{goerlich_resetting_2024}. The acceleration is comparable to that obtained with deterministic shortcut-to-equilibration protocols, establishing resetting as a stochastic shortcut between steady states. A related construction showed that, for a harmonic potential, a spatially dependent resetting rate can be chosen to recover a target Gaussian equilibrium distribution~\cite{Roldan_path_2017}. These examples show that resetting enables controlled state-to-state driving whenever the resetting dynamics can be related to the target distribution.

The same idea is not limited to transitions between equilibrium states. Since resetting naturally generates non-equilibrium steady states, it can also drive transitions between steady states that violate detailed balance~\cite{goerlich_resetting_2024}. This broadens resetting from a shortcut-to-equilibration mechanism to a general strategy for accelerating relaxation toward prescribed steady states.

A complementary result was obtained for resetting within a potential~\cite{sherf_stochastic_2026}. 
For an overdamped particle in a harmonic trap, the dynamics without resetting relaxes thermally to the trap's Boltzmann distribution at intrinsic rate $\omega_0 = \kappa/\gamma$, the ratio of trap stiffness $\kappa$ to bath viscosity $\gamma$.
Adding Poisson resetting at rate $r$ changes both the relaxation pathway and the final state. 
The system no longer equilibrates to the Boltzmann distribution, but instead converges to a resetting-induced non-equilibrium steady state whose shape depends on the resetting protocol~\cite{Pal_diffusion_2016, sherf_stochastic_2026}. 
Sherf et al. \cite{sherf_stochastic_2026} compared the relaxation with and without resetting, defining the relaxation time as the convergence time to the corresponding stationary state via the Kullback-Leibler divergence.
They showed that resetting always reaches its steady state faster than the corresponding thermal process reaches equilibrium. The dynamics are controlled by the dimensionless parameter $S=\frac{\gamma r}{\kappa}=\frac{r}{\omega_0}$, which measures the competition between resetting and trap relaxation. 
For resetting to the trap minimum, the speedup follows the ratio of the second-moment relaxation times, $(S+2)/2$. 

In summary, stochastic resetting accelerates state preparation in two related ways: when the resetting-induced distribution matches an equilibrium target, it acts as a stochastic shortcut to equilibration; otherwise, it still accelerates convergence to a non-equilibrium steady state. In both cases, acceleration comes from stochastically modifying trajectories rather than relying on thermal relaxation.

This acceleration comes at a cost that depends on the implementation. In experiments, resetting requires external intervention (optical traps, chemical driving, or other controls) and generally incurs both thermodynamic and temporal costs~\cite{fuchs_stochastic_2016, pal_integral_2017, gupta_work_2020, Garcia-Valladares_2024}; in simulations, it re-injects configurations or trajectories, so the cost is computational. These trade-offs motivate optimal-control formulations that balance relaxation speed, target-state fidelity, energetic dissipation, temporal overhead, and computational cost~\cite{gupta_stochastic_2021, olsen_thermodynamic_2024, goerlich_time-energy_2026}, discussed further in Sec.~\ref{sec:thermodynamics}.

\subsection{Accelerating rare transition in molecular simulations and kinetics inference}
\label{sec:sr_for_es}

Molecular dynamics (MD) simulations are central to physical chemistry~\cite{blumer_stochastic_2022,Blumer2025review} but face a severe gap between femtosecond integration steps and the long timescales of rare events.
Enhanced-sampling methods such as Metadynamics \cite{laio_metadynamics_2008, Valsson2016, barducci_well-tempered_2008, Barducci2011} address this by biasing the dynamics, but their success relies on correctly identifying the relevant collective variables, itself a severe challenge~\cite{Sidky2020}.
Resetting has recently emerged as a complementary route for enhanced sampling. Since it was recently reviewed by Blumer and Hirshberg \cite{Blumer2025review}, we mention only the key conclusions here.

The practical appeal is that resetting requires no modification of the equations of motion or biasing potential: trajectories are simply terminated at prescribed or random times and restarted from independently sampled initial conditions. Because the segments are statistically independent, resetting is easily parallelizable and can serve as a standalone protocol~\cite{blumer_stochastic_2022} or combined with methods such as Metadynamics~\cite{blumer2024combining}, where restarts cut short unproductive segments and overcome entropic barriers while Metadynamics drives exploration and overcomes enthalpic barriers~\cite{Blumer2025review}.

Beyond acceleration, resetting can also infer kinetic observables of the original, unbiased process~\cite{blumer_stochastic_2022, blumer2024inference, blumer2024short}. This is crucial for molecular simulations, where the goal is often not just to observe transitions more frequently but to estimate the mean FPT without resetting.
To infer the mean first-passage time of the original process from trajectories sampled under Poisson resetting at rate $r^\ast$, one can first use these data to predict the mean first-passage time at higher resetting rates, $r=r^\ast+\Delta r$. This is done through a modified version of Eq.~\ref{eq:MFPT_Poisson_resetting},
\begin{equation}
    \langle T_{r>r^\ast}\rangle =
    \frac{1-\tilde{f}_{T_{r^\ast}}(\Delta r)}
    {\Delta r\,\tilde{f}_{T_{r^\ast}}(\Delta r)} ,
\end{equation}
where $\tilde{f}_{T_{r^\ast}}$ is the Laplace transform of the first-passage-time distribution measured under resetting at rate $r^\ast$. It relies on merging and splitting of Poisson processes remaining Poissonian, so it does not apply to non-Poisson resetting \cite{Ross2019}. The mean first-passage time without resetting is then inferred by extrapolating to $r\to0$, e.g., by polynomial fitting. Generically, one observes a trade-off between speed-up and accuracy, typical of kinetic inference.

More recently, a complementary inference method was developed for resetting at constant time intervals (sharp resetting)~\cite{blumer2024inference, Blumer2025review}. There, the dynamics between resets remain unperturbed, so simulations directly sample the exact short-time first-passage statistics of the unbiased process up to the reset time, while the missing long-time contribution is inferred by fitting the asymptotic tail of the first-passage distribution. This enables inference of non-exponential kinetics, including power-law and multi-exponential behavior, from resetting-accelerated simulations.

\subsection{Resetting to accelerate machine learning}
Although machine learning and molecular dynamics are often viewed as distinct disciplines, both can be formulated as stochastic processes. During neural-network training, model parameters evolve under noisy gradient updates on an effective loss landscape; in generative models such as large language models (LLMs), the stochastic process is the repeated generation of samples from a learned distribution. This raises the question of whether resetting can accelerate machine learning as it does stochastic processes in physical chemistry. Below, we review recent developments in both training and sampling.

\subsubsection{Accelerating training}

\begin{figure}[t!]
    \centering
    \includegraphics[width=\linewidth]{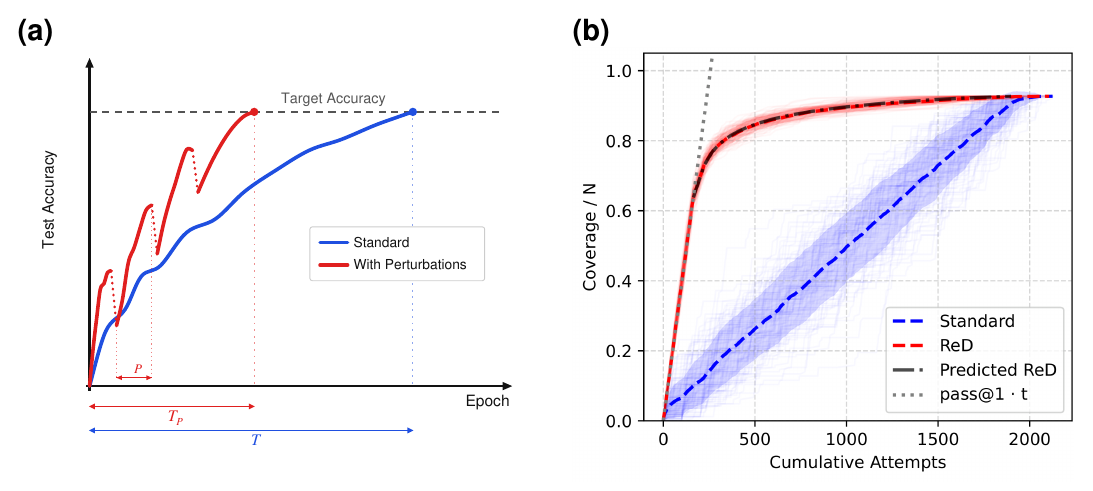}
    \caption{\textbf{Resetting accelerates machine-learning training and sampling.}
  (a) Perturbations, such as full or partial resetting, accelerate convergence to a target accuracy. Schematic illustration of standard training (blue), that reaches the target accuracy only after
  $T$ epochs, whereas training with periodic perturbations, e.g., resetting, applied
  every $P$ epochs (red) reaches the target faster.
  (b) Resetting accelerates LLM sampling: coverage
  (fraction of solved problems, as a function of
  cumulative attempts. Standard sampling (blue) grows slowly, while
  ReD (red), attains near-complete
  coverage with far fewer attempts; the predicted ReD curve (dash-dot)
  closely tracks the observed one, and $\text{pass@}1\cdot t$ (dotted)
  is shown for reference. Shaded bands denote variation across runs. 
  }
    \label{fig:mlfig}
\end{figure}

Model training is sometimes perturbed in ways that amount to variants of resetting. Warm-restart algorithms periodically reset the learning rate to its initial large value~\cite{loshchilov2016sgdr}, and other protocols reset the network parameters to a dynamically updated checkpoint~\cite{bae2024stochastic}. These approaches show resetting can influence learning trajectories, but are usually applied heuristically, without a predictive theory of when a perturbation accelerates training and by how much~\cite{zaidi2023does}.

Meir et al. \cite{meir2025first} recently formulated neural-network training as a first-passage process, defining first-passage as the first time the model reaches a prescribed accuracy (see Fig.~\ref{fig:mlfig}a). This allows perturbations to the training dynamics to be analyzed with stochastic theory, even though the high-dimensional dynamics are not analytically tractable.
To do so, they used a theory by Keidar and Reuveni~\cite{keidar2026universal} for the response of first-passage processes to perturbation, and showed that, when the unperturbed training dynamics settle into a quasi-steady state, measuring the response to a perturbation at a single frequency suffices to predict the response across a broad frequency range. This framework identifies perturbation strategies that accelerate training across datasets, architectures, optimizers, and tasks, establishing a first-passage approach for designing and optimizing training perturbations, including stochastic resetting.

\subsubsection{Accelerating sampling and power-law inference}
Resetting can also accelerate sampling from machine-learning models, particularly LLMs.
Here the parameters are fixed, so the relevant stochastic process is not the training trajectory but the repeated generation of candidate solutions to the same problem.
For verifiable tasks, such as coding or mathematical reasoning, each attempt can be viewed as a stochastic trial, and the relevant first-passage time is the number of attempts required to obtain a correct answer. 

The performance of LLMs is often characterized by pass@$k$, the probability of solving a single problem at least once in $k$ attempts \cite{Kulal2019}. Across many models and datasets, it is found that the large-$k$ behavior of $1-\text{pass@}k$ has an empirical power-law scaling of $k^{-\alpha}$, where $0 < \alpha < 1$ \cite{levi2025, schaeffer2025, kazdan2025}.
In practice, one usually has a fixed budget of attempts, tokens, or compute, and asks how many distinct problems can be solved within it.
This objective, referred to as coverage@cost, naturally leads to a workload-level resetting strategy: rather than spending many attempts on one instance until completion, one occasionally resets by moving to a different unsolved problem and discarding solved ones. 
Meir et al. \cite{meir2026bangbuckimprovinginference} proposed the Reset-and-Discard (ReD) protocol, and showed that resetting always improves coverage@ (see Fig.~\ref{fig:mlfig}b) by reallocating stochastic trials across an ensemble of tasks, and that the optimal resetting strategy resets after every attempt. Moreover, they showed that ReD turns accelerated sampling into a diagnostic tool, enabling efficient estimation of the exponent governing the asymptotic power-law scaling of pass@$k$.

In short, resetting provides a common language for accelerating ML in two regimes: during training, by perturbing trajectories in parameter space, and during sampling, by allocating a global budget across tasks.

 \section{INTERPLAY BETWEEN INFORMATION, FEEDBACK, AND RESETTING}
Until now, we have assumed that resetting is controlled by an external timer $R$, drawn from a resetting-time distribution $f_R(t)$. The same protocol can also be described by a time-dependent resetting rate, that depends only on the elapsed time since the last reset. Denoting $\Psi_R(t)$ as the probability that no reset has occurred up to time $t$, one can write $f_R(t)=\Psi_R(t) r(t)$, where $r(t)$ is the instantaneous resetting rate, conditioned on no earlier resets. This decomposition highlights that we have so far only considered resetting rates that are completely decoupled from the dynamics of the underlying process. A natural extension is: what changes when the reset rate depends on the state of the process, or its history?

It has recently become clear that such state-dependent, or adaptive, resetting strategies can be useful.
For example, state- or history-dependent resetting expands the class of non-equilibrium steady states accessible experimentally \cite{Evans_diffusion_2011_JPhysA, Roldan_path_2017}. In molecular dynamics, avoiding resets near the product region has been shown to expedite simulations further~\cite{Church2025}.
State-dependent resetting also arises naturally, e.g., in an enzyme-substrate complex with multiple states of different off rates \cite{Berezhkovskii_Dependence_2017, Singh_theoretical_2019}, or a chaperone with different affinities to different folding intermediates.

Adaptive resetting rates are difficult to handle because every realization of the underlying process yields a different realization of the resetting rate, so the probability of arriving at a given point at a given time without having been reset depends on the trajectory. Predicting the dynamics then requires summing over all trajectories weighted by their likelihood, a formidable challenge.
Even for a particle diffusing in a potential with space-dependent resetting, this problem is equivalent to a quantum path integral~\cite{Roldan_path_2017}, indicating that analytical solutions are possible only for a handful of cases~\cite{Roldan_path_2017, Tucci_controlling_2020}. In a few ad hoc cases, one can avoid the path integral and obtain the first-passage \cite{Evans_diffusion_2011_JPhysA, Plata_asymmetric_2020, Cantisan_energy_2024, pinsky_diffusive_2020, Ye_random_2022, Singh_theoretical_2019, Berezhkovskii_Dependence_2017, Biswas_target_2025, DelVecchio_Proxitaxis_2026}, steady-state \cite{Evans_diffusion_2011_JPhysA, Plata_asymmetric_2020, DeBruyne_optimization_2020, Ye_random_2022, DeBruyne_resetting_2023}, and thermodynamic properties \cite{Ali_asymmetric_2022, TalFriedman_smart_2025}.

Very recently, Keidar and Blumer et al.\cite{keidar_adaptive_2025} developed a general formulation of adaptive resetting, generalizing all key equations of Sec.~\ref{sec:theory} to any state and time-dependent resetting protocol.
They also showed that a single set of trajectories without resetting, obtained experimentally, numerically, or analytically, suffices to estimate all key observables under adaptive resetting by a simple reweighting: the mean FPT, the entire FPT distribution, the propagator, and the steady-state distribution. Below we briefly review their framework.

\subsection{Steady-states under adaptive resetting}
\label{sec:NESS_with_adaptive}
We define a resetting rate $r(\boldsymbol{x},t)$, where $\boldsymbol{x}$ is the state of the underlying process.
Given a trajectory, $\{ \boldsymbol{x}(t'), 0 \le t' \le t \}$, we can define a random variable describing the resetting time $R$, given this trajectory, via its survival function
\begin{equation}
\Psi_{R | \{\boldsymbol{x}\}}(t) = \exp\left(-\int_{0}^tr(\boldsymbol{x}(t'), t')\,dt'\right). 
\label{probRlet}
\end{equation}
By taking the time derivative on both sides, we get $f_{R | \{\boldsymbol{x}\}}(t) = \Psi_{R | \{\boldsymbol{x}\}}(t) r(\boldsymbol{x}, t)$.
The probability of having the first resetting at time $\tau$ is then just the average over all possible trajectories, $f_R(t)\equiv \left\langle f_{R | \{\boldsymbol{x}\}}(t)  \right\rangle_{ \{ \boldsymbol{x} \} }$.
Extending the renewal Eq. (\ref{eq:propagator_renewal}) to the case of adaptive resetting~\cite{keidar_adaptive_2025}, one gets
\begin{equation}
    p_R(\boldsymbol{x},t)=p_\Psi(\boldsymbol{x},t)+\int_0^tf_R(\tau)p_R(\boldsymbol{x},t-\tau)\,\mathrm{d}\tau.
\end{equation}
Here, $p_\Psi(\boldsymbol{x},t)$ is defined as
\begin{equation}\label{eq:p_psi_def}
    p_\Psi(\boldsymbol{x},t) \equiv
    p(\boldsymbol{x},t)
    \left\langle
    \Psi_{R|\{\boldsymbol{x}^{\prime}\}}(t)
    \,\middle|\,
    \boldsymbol{x}^{\prime}(t)=\boldsymbol{x}
    \right\rangle_{\{\boldsymbol{x}^{\prime}\}} .
\end{equation}
It represents the probability density of finding the process at state $\boldsymbol{x}$ at time $t$, weighted by the probability that no reset has occurred along the trajectory up to that time.
When $\Psi_{R | \{\boldsymbol{x}\}}(t)$ is independent of the trajectory, Eq.~\ref{eq:p_psi_def} reduces to standard resetting, $p(\boldsymbol{x},t) \Psi_R(t)$.
We stress that $p_\Psi(\boldsymbol{x},t)$ and $f_R(t)$ are path integrals whose evaluation is the key challenge in adaptive resetting.

Similarly to Eq.~\ref{eq:propagator_renewal_Laplace}, taking the Laplace transform, we get
\begin{equation}\label{eq:adaptive_Laplace}
    \tilde{p}_R(\boldsymbol{x},s)=\frac{\tilde{p}_\Psi(\boldsymbol{x},s)}{1-\tilde{f}_R(s)}.
\end{equation}
As in Eq.~\ref{eq:poisson_reset_time_evoultion_NESS}, we use the final value theorem and the moment expansion of the Laplace transform $\tilde{f}_R(s) \simeq 1-s\langle R \rangle$, to get the steady state,
\begin{equation}\label{eq:adaptive_NESS}
    p_R^{ss}(\boldsymbol{x})=\frac{1}{\langle R\rangle}\tilde{p}_\Psi(\boldsymbol{x},0)\propto \int_0^{\infty}\mathrm{d}t \, p_\Psi(\boldsymbol{x},t).
\end{equation}
If $p_\Psi(\boldsymbol{x},t)$ or its Laplace transform is available analytically, we obtain the steady state under resetting; but this is almost never the case. Alternatively, trajectories from experiments or simulations can directly estimate $p_\Psi(\boldsymbol{x},t)$ as weighted averages over the data, as we now show.

Consider a set of $N$ trajectories, discretized in time with a time step $\Delta t$. We define the probability that the $i$-th trajectory will be reset at the $j$-th time step given no prior resetting as
\begin{equation}
    \label{eq:pij}
    p^i_j=r(\boldsymbol{x}^i_j,j\Delta t)\Delta t
\end{equation}

Here, $\boldsymbol{x}^i_j$ is the state of the system at the $j$-th time step along the $i$-th trajectory. Then, the probability of not being reset up to time step $j$ following the $i$-th trajectory can be estimated as 
\begin{equation}
\label{eq:psi_i_j}
    \Psi_j^i\approx\prod_{k=1}^{j-1}(1-p_k^i).
\end{equation}
Following the definition of $p_\Psi(\boldsymbol{x},t)$ in Eq.~\ref{eq:p_psi_def}, it can be estimated as,
\begin{equation}
\label{eq:adaptiveres_pred}
    p_\Psi(\boldsymbol{x},j\Delta t)\approx\frac{1}{N}\sum_{i=1}^N \Psi_j^i\delta(\boldsymbol{x}-\boldsymbol{x}_j^i).
\end{equation}
Then, the steady-state distribution is estimated (up to normalization) as a sum of $p_\Psi(\boldsymbol{x},j\Delta t)$ over all time steps $j$, according to Eq.~\ref{eq:adaptive_NESS}.

The strength of this approach is that, from a single set of trajectories without resetting, the properties under adaptive resetting can be predicted by reweighting, letting us scan a large range of protocols and analyze their performance without additional expensive simulations.
For example, Keidar and Blumer et al. \cite{keidar_adaptive_2025} applied this framework to diffusion under adaptive resetting, where the resetting rate depends on position as $r(x)=r_0|x|^\lambda$, with $r_0\geq0$. Two limiting cases were previously solved. For $\lambda=0$, the protocol reduces to diffusion with a constant resetting rate, yielding the Laplace steady state $ p_R^{\rm ss}(x)=\sqrt{\frac{r}{4D}}\,e^{-\sqrt{r/D}|x|}
$ first obtained by Evans and Majumdar \cite{Evans_diffusion_2011}. For $\lambda=2$, the corresponding solution is related to the harmonic oscillator path integral and produces a steady state with Gaussian tails~\cite{Roldan_path_2017}. These examples illustrate the limitation of analytical approaches: different $\lambda$ map onto distinct path-integral problems, making closed-form solutions difficult.

The trajectory-based framework circumvents this limitation. Using a single ensemble of diffusive trajectories generated without resetting, Keidar and Blumer et al. \cite{keidar_adaptive_2025} predicted the resetting-induced steady states for $\lambda\in\{0,1,2,3\}$. This analysis showed that the tails of the steady state follow a stretched-exponential form, $ p_R^{\rm ss}(x)\sim e^{-\nu |x|^{1+\lambda/2}}$, where $\nu$ is a positive constant~\cite{keidar_adaptive_2025}, giving a direct route to these steady states without solving each reset dynamics separately.

\subsection{Acceleration with adaptive resetting}
\label{sec:acc_Adaptive}
A similar analysis can be carried out for first-passage statistics. Weighted averages over trajectories generated without resetting can again be related to the first-passage-time distribution and mean FPT under adaptive resetting.

We start from Eq.~\ref{eq:MFPT_general_resetting}. The first equality in that equation no longer applies, because the resetting time $R$ and first-passage time $T$ are no longer independent. However, the relation, $ \langle T_R\rangle=
    \langle \min(T,R)\rangle/{\Pr(T\leq R)}$, 
still holds, which can be rewritten as
\begin{equation}\label{ISR MFPT TET}
    \langle T_R\rangle=
    \left(\frac{1}{\Pr(T\leq R)}-1\right)
    \langle R \mid R<T\rangle
    + \langle T \mid T\leq R\rangle .
\end{equation}

Keidar and Blumer et al. \cite{keidar_adaptive_2025} showed that all terms on the right-hand side of Eq.~\ref{ISR MFPT TET} can be estimated from an ensemble of $N$ trajectories of the original process, generated without resetting. For such a non-reset trajectory $i$, if the target is reached after $n_i$ time steps, the first-passage time is $n_i\Delta t$. Using $p_j^i$ from Eq.~\ref{eq:pij}, and $\Psi_j^i$ from Eq.~\ref{eq:psi_i_j}, one obtains
\begin{equation}\label{success probability}
    \begin{split}
        \Pr(T\leq R)&\approx\frac{1}{N}\sum_{i=1}^N \Psi^i_{n_i},
    \end{split}
\end{equation}

\begin{equation}\label{cMFPT}
    \begin{split}
        \langle T|T\leq R\rangle &\approx\frac{1}{N\Pr(T\leq R)}\sum_{i=1}^N \Psi^i_{n_i}n_i\Delta t,
    \end{split}
\end{equation}
and
\begin{equation}\label{mean R}
    \langle R|R< T\rangle\approx\frac{1}{N(1-\Pr(T\leq R))}
    \sum_{i=1}^N\sum_{j=1}^{n_i-1}\Psi_{j}^i\,p_j^i\,j\Delta t.
\end{equation}

These equations can identify adaptive protocols that accelerate processes more than regular resetting, by exploiting information on the process's progress and avoiding resets close to the target, as demonstrated on simulations of the protein chignolin in explicit solvent~\cite{Church2025}.
Adaptive resetting can also go beyond scanning protocols: representing the resetting rate with a flexible functional form, such as a neural network, lets us automatically optimize protocols that accelerate first-passage kinetics or generate prescribed non-equilibrium steady states~\cite{keidar_adaptive_2025}, discussed in detail in Sec.~\ref{subsec:Learning of Adaptive Protocols}.

\subsection{Resetting with environmental feedback}

\begin{figure}[ht!]
    \centering
    \includegraphics[width=0.75\linewidth]{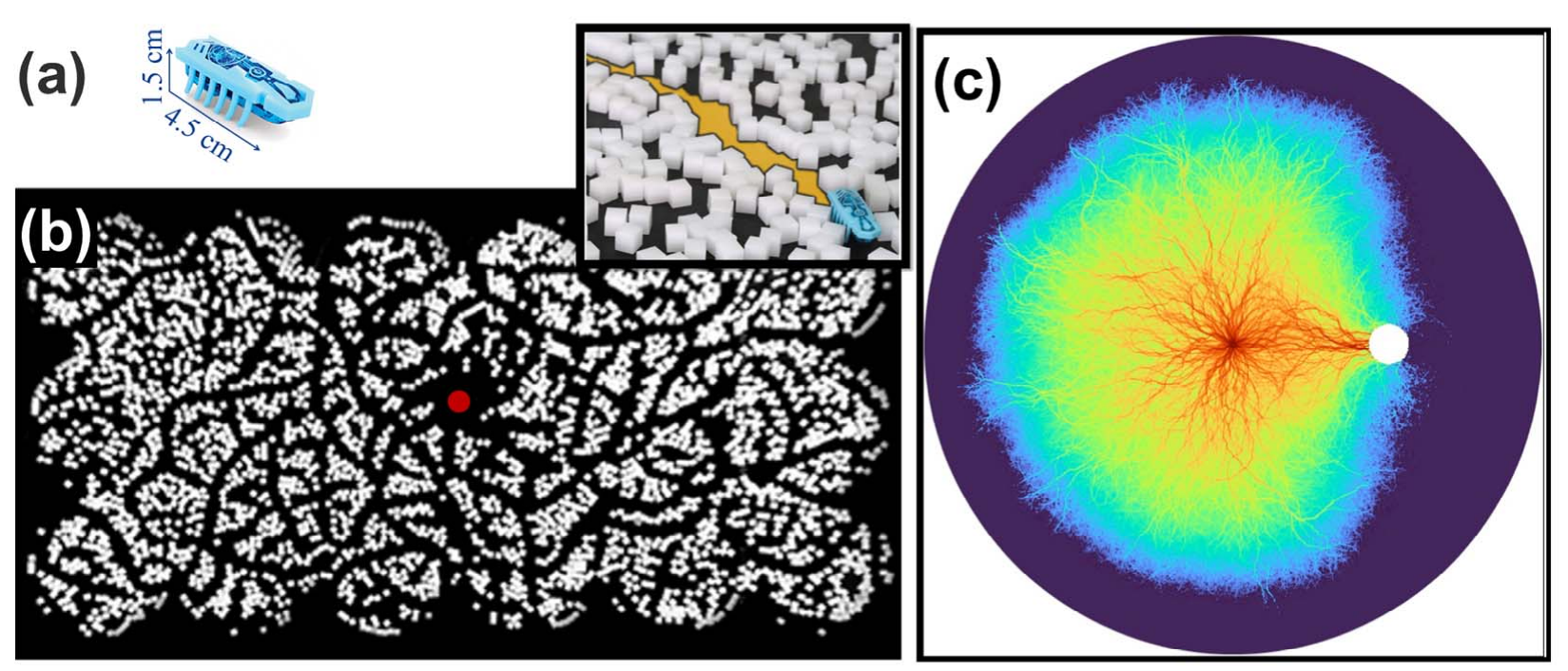}
    \caption{\textbf{Resetting with environmental memory.} (a) A bristle robot (Hexbugs$^{\text{TM}}$ nano). (b) The bristle robot is placed in an arena filled with movable plastic cubes. The bristle robot is reset to the origin (red circle) at a different angle every $20s$, or whenever it reaches the boundary. Its motion through the arena creates obstacle free trails (see inset). At steady state, the arena is riddled with trails, which allows the robot to move faster, resulting in a higher effective diffusion coefficient and decreased mean FPT \cite{altshuler_environmental_2024}, (b) the accumulated trajectories of a chemotactic active Brownian walker. The walker increases its scent secretion upon reaching the target, while the mean FPT is not generally improved by trails formed via chemotactic motion, the target hit rate increases \cite{rudyak_channel_2025}.}
    \label{fig:Env}
\end{figure}

Resetting with environmental feedback may depart from standard renewal descriptions (Secs.~\ref{sec:theory} and \ref{sec:NESS_with_adaptive}-\ref{sec:acc_Adaptive}), because a reset may renew both the searcher and its environment, or only the searcher. In the latter case, information about previous trajectories persists as structural changes to the medium, chemical gradients, depletion fields, or slow environmental degrees of freedom~\cite{altshuler_environmental_2024, rudyak_channel_2025, santra_viscoelastic_2026}. This distinguishes environmental feedback from adaptive resetting, where information directly modulates the resetting rate.

A minimal theoretical realization was recently studied for a probe particle in a viscoelastic bath~\cite{santra_viscoelastic_2026}. The probe is coupled to an auxiliary bath coordinate representing the slow relaxation of the medium; resetting acts only on the probe, while the bath coordinate keeps evolving, so each reset renews the observable coordinate but not the full tracer-bath state. Using exact moment equations, limiting analytical forms, and simulations, Santra et al. showed that this retained bath memory changes both the relaxation and the steady state. For instantaneous resets, strong bath memory produces a two-step relaxation of the probe variance and stationary distributions with Gaussian rather than exponential tails. For finite-time returns, the stationary distribution becomes sensitive to the return velocity, in contrast to e.g., overdamped Brownian motion, which belongs to a class of processes where constant-velocity return protocols leave the stationary density invariant \cite{Pal_TimeDependent_2019,pal2019invariants}. This provides a clean example in which environmental memory changes the effect of resetting on the observed dynamics.

A direct experimental example of environmental memory is a self-propelled active particle (Hexbugs$^{\text{TM}}$ nano) moving through a field of movable obstacles. Under periodic returns to a home position, mechanical collisions reorganize the obstacle landscape and carve persistent physical trails (Fig.~\ref{fig:Env}(a))~\cite{altshuler_environmental_2024}. Because the reset restores the particle position but leaves these structural modifications intact, subsequent trajectories are guided by the trails generated during earlier excursions. This externalized memory breaks renewal at the level of the coupled system, increasing the effective diffusion coefficient and accelerating target acquisition relative to control protocols in which the obstacle configuration is scrambled after each reset~\cite{altshuler_environmental_2024}.

A related mechanism appears in chemical environments. For chemotactic active Brownian particles searching for a consumable target, the interplay between directional bias and nutrient consumption generates self-reinforcing channels in the concentration field ~\cite{rudyak_channel_2025,RUDYAK2026110285}. These channels encode historical exploration and feed it back into subsequent motion, creating a funnel toward the target (Fig.~\ref{fig:Env}(b)), reducing the sensitivity of search performance to target dimensions while increasing the total utilization time of the resource~\cite{rudyak_channel_2025}. Similar feedback principles apply to responsive porous media, where the repeated passage of a searcher or tracer modifies local permeability, connectivity, or accessible pathways.

Environmental feedback therefore reframes information use in stochastic resetting: although the reset rule remains simple and memoryless, the medium registers past trajectories and biases future motion. The search can no longer be modeled by independent renewal intervals, and its efficiency reflects the joint, nonlinear dynamics of resetting, spatial exploration, and environmental memory.

\section{DESIGNING STEADY STATES AND FIRST-PASSAGE TIMES WITH RESETTING}

Generating prescribed stationary distributions is central to many applications in physical chemistry, including preparing target ensembles for thermodynamic pathways, constructing controlled initial conditions for relaxation experiments~\cite{sartor_microarray_2004}, and exploring multidimensional free-energy landscapes~\cite{torrie_nonphysical_1977, laio_metadynamics_2008}. Conventional ensemble design reshapes the potential energy landscape or modifies macroscopic control parameters such as temperature, thereby changing the equilibrium Boltzmann distribution. Stochastic resetting provides a distinct non-equilibrium route: by controlling the reset location and rate, one can shape the stationary distribution and sustain probability currents that are absent from equilibrium ensembles~\cite{goerlich_resetting_2024, sherf_stochastic_2026}.

\subsection{Partial resetting}

The design flexibility of stochastic resetting increases further when reset events do not fully reinitialize the system. In partial stochastic resetting \cite{tal-friedman_diffusion_2022,pierce2021stochastic}, the coordinate is reduced by a fixed fraction rather than returned to a fixed point, such that
\[
x \to ax, \qquad 0 \le a \le 1 .
\]
For a freely diffusing particle under Poissonian partial resetting, the resulting non-equilibrium steady state can be represented as an infinite sum of independent, non-identically distributed, Laplace random variables~\cite{tal-friedman_diffusion_2022}.

The attenuation parameter $a$ provides a continuous control over the shape of the stationary distribution. At $a=0$, the process reduces to total resetting and produces the familiar localized Laplace profile~\cite{Evans_diffusion_2011}. As $a \to 1$, the effect of each reset becomes progressively weaker, and the stationary distribution crosses over to a Gaussian-like profile~\cite{tal-friedman_diffusion_2022}. Similar behavior persists for diffusion with drift. Partial resetting therefore provides a simple yet powerful design parameter: tuning the attenuation parameter continuously interpolates between Laplace- and Gaussian-like stationary distributions, mimicking the confinement produced by different potential landscapes without modifying the underlying dynamics.

This design perspective extends beyond the steady-state profiles discussed above. A general framework for time-dependent partial resetting has been developed for homogeneous Markov processes, yielding exact expressions for the time-dependent propagator and, through its long-time limit, the stationary distribution~\cite{di2023time}. More general partial resetting protocols may further expand the range of attainable steady states. For example, refractory periods produce stationary mixtures of exploration and refractory subpopulations~\cite{olsen_refractory_2024}. Allowing both positive and negative rescaling factors, $-1 \le a \le 1$, leaves the stationary distribution unchanged but dramatically alters first-passage behavior, with negative rescaling accelerating target search relative to conventional resetting~\cite{biroli2024partialresetting}. 

\subsection{Learning of adaptive protocols}
\label{subsec:Learning of Adaptive Protocols}

Adaptive protocols dramatically extend the flexibility of resetting, and raise an interesting question: Can we design adaptive protocols to tailor the properties of the process with resetting? One would like to represent the resetting rate as a flexible function with many parameters, and find the combination of parameters that minimizes some cost. This cost can be the mean FPT with resetting, if we want to expedite kinetics, or the distance to a desired steady-state. 

\begin{figure}[ht!]
    \centering
    \includegraphics[width=0.75\linewidth]{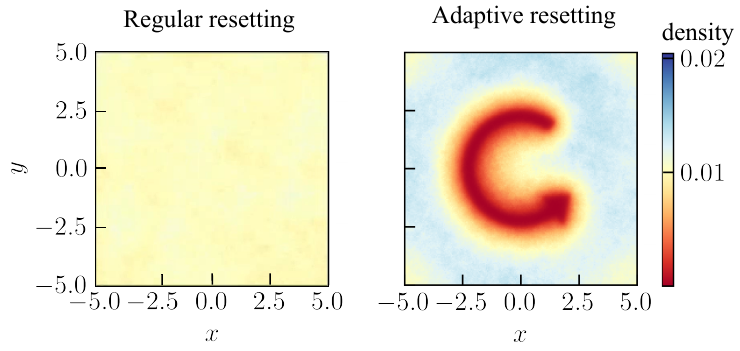}
    \caption{\textbf{Designing steady-states with adaptive resetting.} State-independent resetting to a uniform distribution of diffusing particles in a two dimensional box leads to a uniform steady-state (left). Adaptive resetting of the same particles can break this uniformity, leading to complex steady-states.}
    \label{fig:adapritve_ss_design}
\end{figure}

Keidar and Blumer et al. \cite{keidar_adaptive_2025} first demonstrated that adaptive resetting can predict and design complex non-equilibrium steady states by tailoring the spatial dependence of the resetting rate, reaching target distributions inaccessible with state-independent protocols (Fig.~\ref{fig:adapritve_ss_design}).
They then went beyond steady states and tailored an adaptive rate to lower mean FPTs by representing $r(\boldsymbol{x},t)$ as a neural network. At each training step, the adaptive-resetting equations of Sec.~\ref{sec:acc_Adaptive} predicted the desired observable from trajectories without resetting, providing a differentiable loss for stochastic gradient descent. Using the mean FPT as the loss, they optimized the protocol and accelerated conformational transitions of chignolin. Although the machine-learning implementation has so far been demonstrated only for mean FPT optimization, the same framework can optimize any observable predicted by the adaptive-resetting theory, including the full first-passage-time distribution, the propagator, and prescribed steady-state distributions.

Muñoz-Gil et al. \cite{Munoz-Gil_2025} showed that adaptive resetting protocols can also be optimized using reinforcement learning. Unlike the reweighting-based framework of Keidar and Blumer et al. \cite{keidar_adaptive_2025}, this approach does not exploit the renewal structure and therefore requires trajectories with resetting throughout training. Training is thus typically more computationally demanding and, in experimental settings, requires automated data acquisition, but it naturally accommodates the simultaneous optimization of resetting and additional control parameters.

\section{RESETTING IN MANY BODY SYSTEMS}

Stochastic resetting in interacting particle systems has been reviewed in Ref.~\cite{nagar_stochastic_2023}. Here, we focus on developments most relevant to this review: how resetting protocols generate correlations, modify search kinetics, and reshape non-equilibrium steady states in systems with many degrees of freedom.

The renewal structure depends on what is reset. A molecule, polymer, or colloidal assembly is a single system with many coupled degrees of freedom. If all are reset together, the process is renewed in the full state space and the renewal framework still applies. The same holds for global resetting of an interacting many-particle configuration: the propagator without resetting and the resetting-time distribution suffice to infer the dynamics under resetting, as demonstrated in~\cite{vatash_numerical_2025}. By contrast, in local or batch resetting, only one particle or a subset is reset while the rest of the coupled system remains unchanged~\cite{miron_diffusion_2021}, so the reset does not renew the full many-body dynamics, and the unreset degrees of freedom can carry memory across reset events.

\begin{figure}[t!]
    \centering    \includegraphics[width=0.75\linewidth]{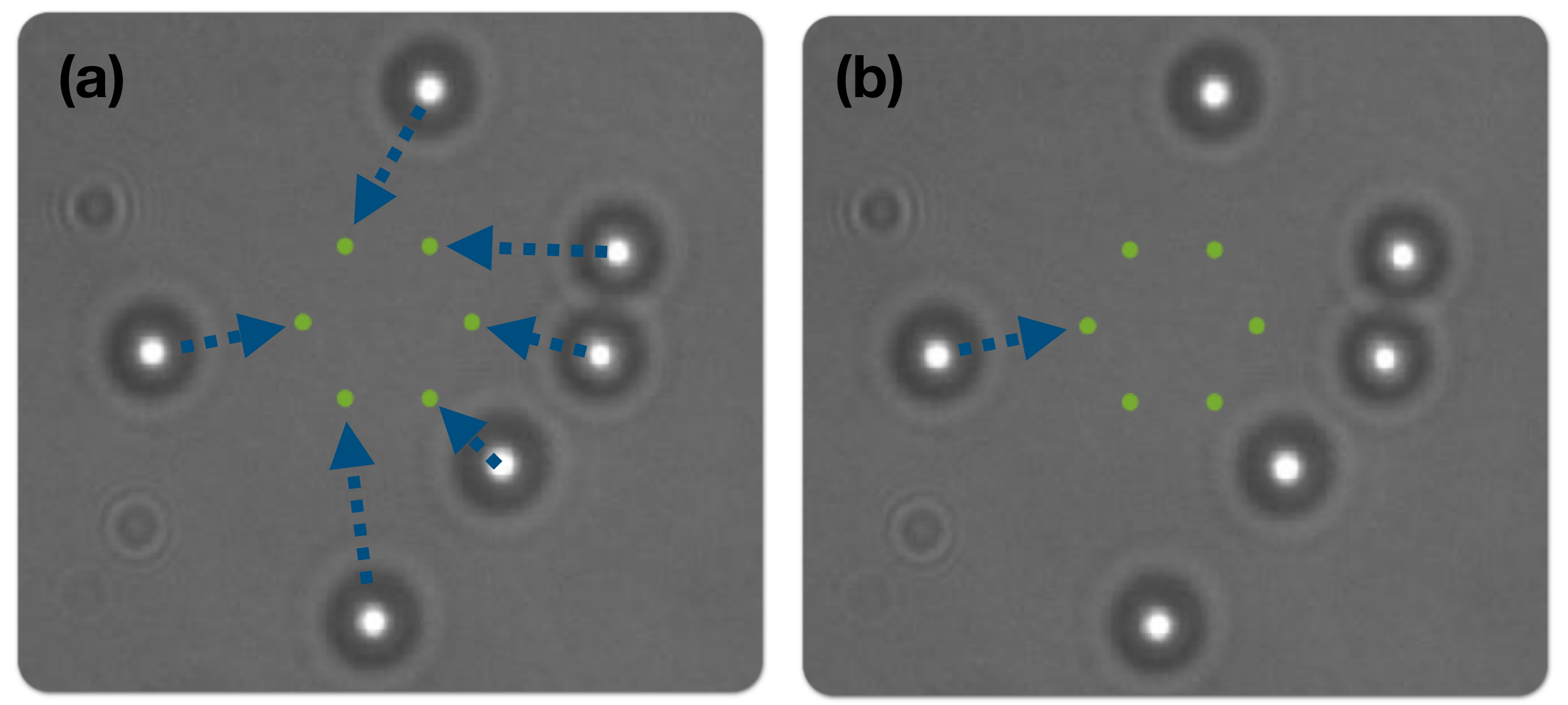}
    \caption{\textbf{Comparison of resetting protocols for colloidal particles trapped in a hexagonal array \cite{vatash_manybody_2025}.} (a) Global resetting, where all particles are returned simultaneously, contrasted with (b) Local resetting, where individual particles are reset to the closest trap. Blue arrows denote the resetting trajectories.}
    \label{fig:Global_local}
\end{figure}

Even without direct interactions, resetting can generate correlations. In simultaneous global resetting, illustrated in Fig.~\ref{fig:Global_local}(a), all particles share a common reset clock, so their positions are statistically coupled through the elapsed time since the last reset~\cite{biroli_extreme_2023, demauro_nonpoissonian_2026}. In threshold, also known as first-passage, resetting~\cite{DeBruyne_optimization_2020} independent searchers are collectively returned to the origin whenever any one of them reaches a prescribed boundary, externally coupling their search trajectories and modifying first-passage statistics~\cite{Biswas_target_2025, biroli_first_2026}. In batch resetting, illustrated in Fig.~\ref{fig:Global_local}(b), only a random subset of $m$ particles from a population of size $N$ is reset at each event~\cite{demauro_batch_2026}. Single-particle observables are then affected mainly through a rate rescaling, whereas two-particle correlations depend non-trivially on $m/N$, allowing protocol-level tuning of emergent spatial correlations~\cite{demauro_batch_2026}.

Interactions add a second source of many-body structure. For diffusing particles with excluded-volume interactions on a one-dimensional lattice, local resetting to the origin produces density profiles governed by the interplay between reset attempts, particle blocking, and collective relaxation~\cite{miron_diffusion_2021}. In colloidal systems, finite-time resetting competes with steric repulsion, hydrodynamic coupling, and density relaxation; resetting particles into a localized region compacts the suspension and reshapes the stationary density and current profiles near the reset site~\cite{vatash_manybody_2025}. Density-level theories and exactly solvable models, including global density equations and Dyson Brownian motion under resetting, connect microscopic reset rules to collective fluctuations and many-body steady-state structure~\cite{bressloff_global_2024,biroli_dyson_2025}.

The key distinction in many-body resetting is whether resetting renews the entire system or only part of it. Global resetting preserves the predictive power of the renewal framework, enabling analysis and design of many-body steady states, first-passage properties, and interparticle correlations. Local or batch resetting, in contrast, creates correlations and memory beyond renewal theory, offering new opportunities to engineer collective non-equilibrium behavior while posing new theoretical challenges.

\section{THERMODYNAMICS OF RESETTING}
\label{sec:thermodynamics}

\subsection{The cost of maintaining a resetting steady state}
\label{sec:cost_ss}
In many physical implementations, resetting breaks detailed balance by displacing the system from its current state to a prescribed reference state or region. Stochastic thermodynamics provides the natural framework for quantifying the energetic and entropic consequences of this operation~\cite{fuchs_stochastic_2016, pal_integral_2017}. Resetting generally maintains a non-equilibrium steady state with finite entropy production, whose magnitude depends on the distribution of reset positions and on the physical protocol used to return the system~\cite{Mori2023}. For idealized instantaneous resetting, the work distribution reflects the positional statistics at reset times~\cite{gupta_work_2020, Sunil2023}. More generally, resetting can be viewed as an information-erasing operation: each reset removes information about the current state of the system, leading to an irreducible contribution bounded by the Landauer principle~\cite{Landauer1961Irreversibility, Berut2012Experimental}, while finite-time returns dissipate additional work in a protocol-dependent manner~\cite{olsen_thermodynamic_2024, gupta_thermodynamic_2025}.

These costs have been quantified in colloidal experiments. In one implementation, a freely diffusing Brownian particle is reset by a holographic optical tweezer that tracks and drags the particle back to the origin at constant velocity~\cite{tal-friedman_experimental_2020}. The energy per reset is set by the optical power and the return time, and trap stability imposes a maximum return velocity, hence a minimum energy per reset. The measured cost is orders of magnitude larger than $k_{\rm B}T$, underscoring that efficient physical resetting remains an experimental challenge. A complementary implementation resets a particle in a weak harmonic trap by rapidly changing the trap stiffness~\cite{Goerlich2025}. In this case, slow quenches can reduce the work per reset, but even optimized quasistatic protocols retain a finite minimal cost. These experiments make explicit the connection between resetting, Maxwell-demon-like feedback, information erasure, and the Landauer bound~\cite{fuchs_stochastic_2016, gupta_thermodynamic_2025}.

In simulations, resetting is often implemented algorithmically by reassigning coordinates or reloading saved configurations, so the mechanical work of a physical reset is not part of the modeled dynamics. Algorithmic resetting may thus carry no energetic cost within the model, although interpreting the reset as a thermodynamic operation still entails information erasure and logical irreversibility.

\subsection{The cost of a search}

During a search, resetting can reduce the mean first-passage time, but only by adding reset events that must be implemented physically through some return protocol, which often carries a temporal and energetic cost~\cite{Pal2020Search, Sunil2023}. In colloidal diffusion experiments with optical-tweezer resetting, this cost was measured directly in first-passage settings, showing that physical search acceleration by resetting carries a finite energetic cost~\cite{tal-friedman_experimental_2020}. For overdamped Brownian search with finite-time returns, this produces a speed-cost trade-off: resetting can accelerate search over the range of rates where it is beneficial, but increasing the number of returns also increases the accumulated work. Related bounds can be formulated using thermodynamic uncertainty relations for first-passage processes~\cite{Pal2021, Pal2023}, and varying the return protocol defines a Pareto front in the plane of mean first-passage time versus energetic cost~\cite{Singh2024, DeBruyne_resetting_2023}.

As an example, for an overdamped Brownian particle, Goerlich et al. ~\cite{goerlich_time-energy_2026} used optimal transport considerations to design a return protocol with duration $t_{\rm rst}$. In this setting, the average work associated with the return obeys
\begin{equation}
    \langle W_{\rm OT}^{\rm rst}\rangle =
    \frac{2\alpha k_{\rm B}T}{r\,t_{\rm rst}},
    \label{eq:OT_cost}
\end{equation}
where $\alpha$ is a dimensionless factor that encodes the mean squared return distance of the trajectories that are reset, measured in units of the diffusive length scale between resets~\cite {goerlich_time-energy_2026}. This expression is a model-specific lower bound for finite-time overdamped return protocols, not a universal cost of resetting. Inertial, active, viscoelastic, and many-body implementations can have different energetic and temporal costs.

Adaptive resetting provides another way to improve the trade-off. By conditioning the reset decision on the particle's position and avoiding resets when it is already near the target, smart resetting reduces the energetic cost of return relative to Poisson resetting at comparable search performance~\cite{TalFriedman_smart_2025}. Such feedback does not eliminate thermodynamic constraints; it shifts part of the accounting to information. First-passage-triggered information engines show that feedback conditioned on a first-passage event carries an information term, determined by the first-passage-time distribution, which must be included in the energetic balance~\cite{Bellon2025}.

\subsection{The cost of swift equilibration}
\label{sec:cost_swift}

The thermodynamic cost of resetting takes on a distinct character when resetting is
used not to locate a target but to drive a system between prescribed distributions, as
discussed in Sec.~\ref{sec:StateToState}. In this context, the relevant figure of merit
is the work required to complete a controlled state-to-state transition within a given duration. As noted there, resetting-based swift equilibration achieves comparable speedups to deterministic shortcut-to-equilibration protocols, but through a fundamentally different mechanism: the deterministic protocols continuously deform an external potential and require precise knowledge of the instantaneous probability density, whereas resetting interrupts the dynamics stochastically and requires only a reset rate and position~\cite{goerlich_resetting_2024}. Their thermodynamic costs are similarly comparable: both pay a work penalty above the
quasistatic limit that grows as the protocol duration is reduced, and both are bounded
below by the same time-energy trade-off~\cite{Pires2023, goerlich_consistent_2026}. 
The practical advantage of
resetting lies in its robustness: unlike ESE, it does not require knowledge of the
potential landscape and remains effective in anharmonic and more complex settings where
deterministic shortcuts become difficult to construct~\cite{goerlich_resetting_2024}.

\section{CONCLUSIONS AND OUTLOOK}
Stochastic resetting has developed from a minimal model of search acceleration into a broad framework for controlling non-equilibrium dynamics. Its two central consequences, the creation of non-equilibrium steady states and the modification of first-passage kinetics, are now relevant across physical chemistry. In biophysics, resetting concepts clarify how molecular and enzymatic processes can be regulated by restart-like events; in computational chemistry, it provides new routes for enhanced sampling and for inferring rare-event kinetics; in soft-matter experiments, it offers a controllable way to drive colloidal systems away from equilibrium. Throughout this review, we have emphasized results connecting these developments, showing that the same renewal framework that predicts the effect of resetting also allows the properties of the original process to be inferred from the accelerated one. Resetting is therefore useful in three complementary ways: as a theoretical tool for analyzing non-equilibrium and biological processes, as an experimental protocol for engineering steady states and controlled relaxation pathways, and as a computational strategy for accelerating simulations and machine learning while extracting kinetic observables.

Looking forward, we highlight three frontiers we find particularly exciting.
The first is theoretical. The widespread applicability of resetting stems from its renewal framework, which allows the same equations to describe an enzyme, a diffusing colloid, a molecular dynamics trajectory, and the training of a neural network. Yet many of the recent developments reviewed here, e.g., partial resetting, many-body resetting, and environmental memory, break renewal in different ways. For these, existing results remain mostly case-specific, and no comparable unifying framework has yet replaced the role played by renewal theory. Identifying what survives when renewal is lost, and whether a common structure underlies these disparate extensions, is one of the central theoretical challenges in the field.

The second frontier is experimental. Resetting has been realized in macroscopic robotic systems and in colloidal systems, where external feedback or optical control can return the system to a prescribed state. 
At the molecular scale, the first challenge is to identify chemical reactions that might benefit from resetting, and what the control parameters are. Then, developing experimental methods to observe and manipulate resetting in molecular systems, for example, by optically or chemically returning them to prescribed states, would not only enable designed resetting protocols, but also allow naturally occurring restart events in enzymes, molecular chaperones, and other systems to be directly observed and controlled.

The third frontier is computational. Can we design resetting protocols that not only lead to desired first-passage or steady-state properties but also balance other observables, such as the thermodynamic cost? 
Adaptive resetting offers a framework for stochastic control. The missing ingredient is the thermodynamic cost. If we can estimate the work associated with an adaptive protocol represented by a neural network from trajectories without resetting, we could optimize energy–time trade-offs automatically, without ever running the reset dynamics.

\bibliographystyle{ar-style3}
\bibliography{literature}

\end{document}